\documentclass[prl,reprint,showpacs,floatfix,aps
twocolumn,
]{revtex4-1}
\usepackage{amsfonts}
\usepackage{amsmath,amssymb}
\usepackage{cancel,verbatim}
\usepackage{graphicx,float}
\usepackage{color}
\usepackage{bm,subfigure}

\usepackage{epsfig}

\newcommand{\beq}{\begin{equation}}  
\newcommand{\eeq}{\end{equation}}  
\newcommand{\bea}{\begin{eqnarray}}  
\newcommand{\eea}{\end{eqnarray}}  
\newcommand{\nueff}{\nu_{\rm eff}}
\newcommand{\fnueff}{\tilde{\nu}_{\rm eff}}
\newcommand{\eps}{\varepsilon}
\renewcommand{\vec}[1]{\bm{#1}}
\newcommand{\fvec}[1]{\tilde{\bm{#1}}}
\newcommand{\fomega}{\tilde{\omega}}

\newcommand{\Ein}{E_{\rm IN}}

\newcommand{\red}[1]{{\textcolor{black}{#1}}}
\newcommand{\blue}[1]{{\textcolor{black}{#1}}}

\begin{document}

\title{Phase Transition to Large Scale Coherent Structures in Two-Dimensional Active Matter Turbulence}
\author{Moritz Linkmann$^1$, Guido Boffetta$^2$, M.~Cristina Marchetti$^{3}$ and Bruno Eckhardt$^1$}
\affiliation{
$^1$Fachbereich Physik, Philipps-Universit\"at of Marburg,  D-35032 Marburg, Germany \\
$^2$Dipartimento di Fisica and INFN, Universit\`a di Torino, via P. Giuria 1, 10125 Torino, Italy \\
$^3$Department of Physics, University of California, Santa Barbara, California, 93106, USA}

\date{\today}

\begin{abstract}
The collective motion of microswimmers in suspensions induce patterns of
vortices on scales that are much larger than the characteristic size of a microswimmer,
attaining a state called bacterial turbulence. Hydrodynamic turbulence acts on even larger
scales and is dominated by inertial transport of energy. Using an
established modification of the Navier-Stokes equation that accounts for the
small-scale forcing of hydrodynamic flow by microswimmers, we study 
the properties of a dense suspension of microswimmers in two dimensions, where
the conservation of enstrophy can drive an inverse cascade through which energy
is accumulated on the largest scales. We find that the dynamical and statistical properties 
of the flow show a sharp transition to the formation of vortices at the largest 
length scale. The results show that 2d bacterial and hydrodynamic turbulence 
are separated by a subcritical phase transition. 
\end{abstract}

\maketitle

Thin layers of bacteria in their planctonic phase form structures that are 
reminiscent of jets and vortices in turbulent flows \cite{Dombrowski04,Dunkel13PRL,Gachelin14}.
This state has been called 
``bacterial turbulence" \cite{Dombrowski04} because of the shape and form of the patterns,
and has been seen in many swimming microorganisms
\cite{Dombrowski04,Dunkel13PRL,Gachelin14} 
and active nematics \cite{Sanchez12,Zhou14,Giomi15}. 
Bacterial turbulence usually appears on scales much smaller than those of
hydrodynamic
turbulence, with its inertial range dynamics and the characteristic energy cascades
\cite{Frisch95}. A measure of this separation is the Reynolds number, which is of order
$10^{-4}-10^{-6}$ for an isolated swimmer in a fluid at rest \cite{Purcell77} and typically
several tens of thousands in hydrodynamic turbulence. 
Recent studies of the rheology
of bacterial suspensions have indicated, however, that the active motion of pusher-type
bacteria can lower considerably the effective viscosity of the suspension
\cite{Hatwalne04,Liverpool06,Sokolov09,Gachelin13,Lopez15,Marchetti15}, to the point where it approaches 
an active-matter
induced ``superfluid'' phase where the energy input from active processes compensates viscous dissipation~\cite{Cates08,Giomi10}. 
In such a situation the collective action of 
microswimmers can produce a dynamics that may 
be influenced by the inertial terms.
In two dimensions, a possible connection to hydrodynamic turbulence is particularly intriguing
because the energy cascade proceeds from small to large scales and can 
result in an accumulation of energy at the largest 
scales admitted by the domain, thereby forming a so-called condensate 
\cite{Kraichnan67,Hossain83,Smith93,Alexakis18}. If bacterial turbulence can couple to 
hydrodynamic turbulence, then the inverse cascade in 2d provides a mechanism
by which even larger scales can be driven. 
We here discuss the conditions under which such a coupling between 
bacterial and hydrodynamic turbulence can occur.

A dense bacterial suspension \red{
consists of
active swimmers in a solvent. It is generally described in terms of coupled
equations for the flow velocity of the suspension and a polarization vector
field  that captures} the coarse-grained dynamics of the microswimmers.
%
\red{In most previous studies the fluid flow was slaved to the swimmer dynamics, so that the equations
focussed on the velocity or polarization of the swimmers~\cite{Wensink12,Srivastava16,Putzig16}. Here, following
recent work by S{\l}omka and Dunkel ~\cite{Slomka15EPJST,Slomka17PNAS}, we
examine instead the effective equation for the fluid flow, obtained 
by slaving the swimmer velocity to the velocity of the suspension \blue{\cite{Linkmann19b}}.} 
Both approaches 
incorporate activity via active stresses that provide a forcing 
in the dynamical equations and yield minimal 
models that capture the pattern-formation process associated with
bacterial turbulence 
\cite{Wensink12,Slomka15EPJST}. 
The effective Navier-Stokes equation for the fluid introduced in Refs.~\cite{Slomka15EPJST,Slomka17PNAS} 
bears a strong similarity to models studied in the context of inertial
turbulence \cite{Beresnev93,Tribelsky96}, \red{hence providing an excellent starting point for examining the 
relation between bacterial and hydrodynamic turbulence.}

These effective models 
have been 
compared with experiments in {\it Bacillus
subtilis} \cite{Wensink12,Dunkel13PRL,Slomka17PNAS}, 
and have been widely used for
investigating  active turbulence 
\cite{Dunkel13NJP,Bratanov15,Slomka15EPJST,Srivastava16,Putzig16,Slomka17PNAS,Slomka17PRF,
Doostmohammadi17,James17,Slomka18JFM}.  In this Letter we \blue{study} 
the
connection between 2d bacterial and hydrodynamic turbulence systematically
within a model that focusses on the \red{suspension} flow and is, 
in that sense, independent of details of the bacterial motion.
We use the model of Refs.~\cite{Slomka15EPJST,Slomka17PNAS}, and 
slightly modify its structure 
so that
that we have a single parameter that controls the strength of the bacterial
forcing. Increasing this parameter, we find a discontinuous  transition to flow
states which are hydrodynamically turbulent in the strict sense, that is, they
display an inverse energy cascade characterized by a scale-independent energy
flux \cite{Kraichnan67,Frisch95,Boffetta14ARFM}. 

For the model \cite{Slomka15EPJST,Slomka17PNAS,Slomka17PRF}, we take
the velocity field $\vec{u}$ to be incompressible,
$\nabla \cdot \vec{u}=0,$
and periodic in a rectangular domain. 
The momentum balance is Navier-Stokes like (with the density scaled to 1),
\beq
\label{eq:momentum}
\partial_t \vec{u}+ (\vec{u}\cdot\nabla) \vec{u} + \nabla p 
= \nabla \cdot {\bm \sigma} \ ,
\eeq
where $p$ is the pressure and ${\bm \sigma}$ the stress tensor.
The stress tensor contains 
three adjustable parameters $\Gamma_i$,
\beq
\label{eq:stress}
\sigma_{ij} = \left(\Gamma_0 - \Gamma_2 \Delta + \Gamma_4 \Delta^2 \right)
 \left(\partial_i u_j + \partial_j u_i\right) \ ,
\eeq
and is most conveniently discussed in Fourier space, where the dissipative term
in Eq.~\eqref{eq:momentum} can be used to introduce 
an effective viscosity, $\nabla \cdot {\bm \sigma} = \nueff \Delta \vec{u}$,
with
\beq
\fnueff(k) = \Gamma_0 + \Gamma_2 k^2 + \Gamma_4 k^4 \ ,
\label{eq:stress_fourier}
\eeq
where $\fnueff$ is the Fourier transform of $\nueff$.
The first parameter $\Gamma_0$ corresponds to the kinematic viscosity, and $\Gamma_4>0$ 
ensures that modes with large $k$ are always damped by hyperviscosity.
If $\Gamma_2<0$ and sufficiently negative, 
the effective viscosity becomes negative in a 
range of wave numbers, thus providing a source of energy and instability. 
This is the only forcing in the model, 
it corresponds to the mesoscale vortices observed in bacterial 
turbulence, and without it all fields decay.
In 2d and for a suitable set of parameters statistically
stationary states with an inverse energy transfer from the band of forced wave numbers
to smaller wave numbers have been found in variants of both minimal models 
\cite{Slomka15EPJST,Oza2016,Mickelin2018}. 
The energy spectrum in Ref.~\cite{Slomka15EPJST} showed a scaling exponent 
close to the Kolmogorov value of $-5/3$, 
characteristic of the constant-flux inverse energy cascade in 2d turbulence 
\cite{Boffetta14ARFM}. Small condensates were observed in Refs.~\cite{Oza2016,Mickelin2018}. 

In the stress model given by Eq.~\eqref{eq:stress_fourier} $\Gamma_2$ determines
not only the strength but also the range of wave numbers that are forced.  In
order to eliminate this influence, we introduce a variant of the model, where
the bacterial forcing is modeled by a piecewise constant viscosity (PCV) in
Fourier space. We take
\beq
\label{eq:nu_k}
\tilde\nu(k) = 
\begin{cases}
\nu_0 > 0 \quad \text{for} \quad k < k_{\rm min} \ ,\\
\nu_1 < 0 \quad \text{for} \quad k_{\rm min} \leqslant k \leqslant k_{\rm max} \ ,\\
\nu_2 > 0 \quad \text{for} \quad k > k_{\rm max} \ .
\end{cases}
\eeq
where $\nu_0$, like $\Gamma_0$, is the kinematic viscosity
of the \red{suspension} and $\nu_1$ and $\nu_2 > \nu_0$ 
correspond to higher-order terms $\Gamma_2 k^2$ and $\Gamma_4 k^4$, 
respectively, in the gradient expansion of the active stresses in Eq.~\eqref{eq:stress_fourier}. 
That is, as in the model of Refs.~\cite{Slomka17PNAS,Slomka15EPJST}, 
they arise from a linear relation between the \red{suspension} flow and the bacterial forcing.
\blue{The latter is justified for dense 2d suspensions, where both polarization       
and suspension velocity are solenoidal, through a reduction in degrees of freedom \cite{Linkmann19b}.}
With this model, the forced wavenumbers are confined to the interval 
$[k_{\rm min},k_{\rm max}]$ and the strength of the forcing is controlled by $\nu_1 < 0$. 
In what follows we carry out a 
parameter study of the PCV model where $\nu_1$ is the only variable parameter. 
   
We solve the 2d PCV model in vorticity formulation 
\beq
\label{eq:pcv-spectral}
\partial_t \fomega(\vec{k}) + 
\mathcal{F}_{\vec{k}}{\left[\vec{u} \cdot \nabla\omega\right]}
= -
\tilde\nu(k) k^2 \fomega(\vec{k})\ ,
\eeq
where  $\omega$ is the vorticity, $\omega(x_1,x_2)\vec{e}_3 = \nabla \times
\vec{u}(x_1,x_2)$, and $\tilde{\cdot}\equiv\mathcal{F}_{\vec{k}}[\cdot]$
denotes the Fourier transform.  The equations are integrated in Fourier space,
in a domain $[0,2\pi]^2$ with periodic boundary conditions, and using the
standard pseudospectral method with full dealiasing according to the 2/3rds
rule \cite{Orszag71}.  The simulations are run without additional large-scale
dissipation terms, until a statistically stationary state is reached. 
As this can take a long
time, we used a resolution of $256^2$  collocation points to 
explore the parameter space, and confirmed the results  
for a few isolated parameter values with higher resolution,
see Table~\ref{tab:simulations-pcv}.  Different resolutions can be
mapped onto each other using the invariance of Eq.~\eqref{eq:pcv-spectral}
under the transformation
\begin{align}
\label{eq:scaling}
x \to \lambda x, \ \ 
t \to t, \ \ 
\nu \to \lambda^2 \nu, \ \ 
\vec{u} \to \lambda \vec{u}, \ \ 
\omega \to \omega. 
\end{align}
For all simulations the initial data are Gaussian-distributed random velocity
fields. 

\begin{table}[t]
\centering
\begin{tabular}{cccccccc}
\hline
$N$ & $|\nu_1/\nu_0|$ & $\nu_2/\nu_0$ & $k_{\rm min}$ & $k_{\rm max}$ & Re & $U$ & $L$ \\
\hline
256 & 0.25-7.0 & 10.0 & 33 & 40 & 19-13677 & 0.29-7.77 & 0.07-1.92 \\
1024 & 1.0 & 10.0 & 129 & 160 & 45 & 0.027 & 0.029 \\
1024 & 2.0 & 10.0 & 129 & 160 & 226 & 0.041 & 0.094 \\
1024 & 5.0 & 10.0 & 129 & 160 & 132914 & 1.17 & 1.93 \\
\hline
\end{tabular}
 \caption{
Parameters used in DNSs of the piecewise constant viscosity model and resulting 
observables. The number of grid points in each coordinate is denoted by $N$,  
the viscosity $\nu_0$ 
and $\nu_1$, $\nu_2$, $k_{\rm min}$, $k_{\rm max}$ are the parameters in Eq.~\eqref{eq:nu_k}. The Reynolds number 
${\rm Re} = UL/\nu_0$ is based on $\nu_0$, the root-mean-square velocity $U$
and the integral length scale $L = 2/U^2 \int_0^\infty dk \ E(k)/k$, 
with $\nu_0 = 1.1 \times 10^{-3}$  for $N = 256$ and 
$\nu_0 = 1.7 \times 10^{-5}$  for $N = 1024$.
Averages in the statistically stationary state are computed from at least
1800 snapshots separated by one large-eddy turnover time $T=L/U$.
 }
 \label{tab:simulations-pcv}
\end{table}

\begin{figure}[t]
\centering
	\includegraphics[width=\columnwidth]{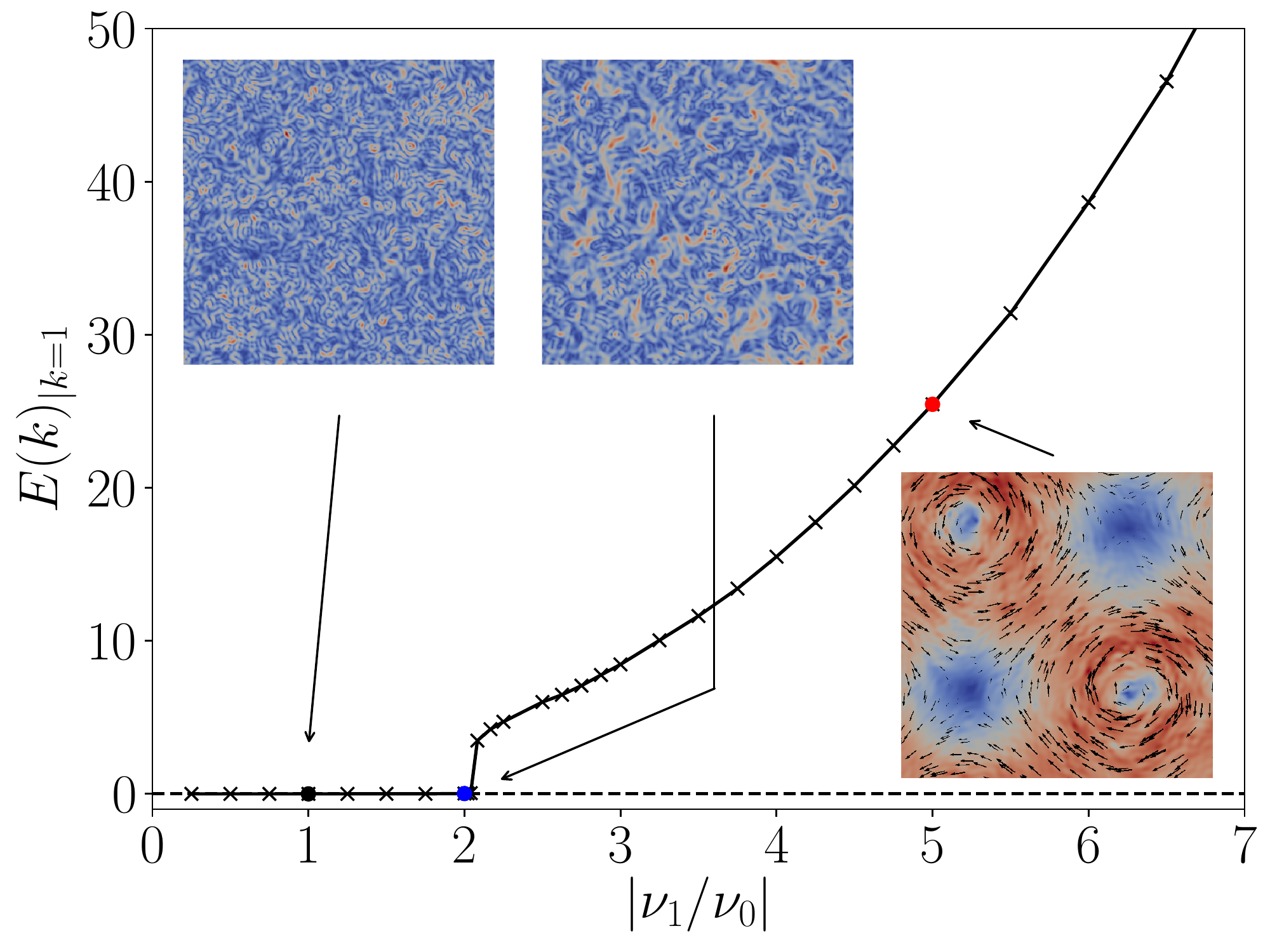}
\vspace{-2em}
 \caption{
         (Color online) Mean energy at the largest scale as a function
         of $\nu_1$. 
         The black, blue and red dots correspond to cases 
         $|\nu_1/\nu_0| = 1$ , $|\nu_1/\nu_0| = 2$ and $|\nu_1/\nu_0| = 5$, respectively, and the corresponding visualisations show $|\vec{u}(\vec{x})|$.
         }
 \label{fig:E1_vs_nu1_caseA}
\end{figure}

A measure of the formation of large scale structures is the energy at the largest scale, 
$E_1 \equiv E(k=1)$, 
where
\beq
\label{eq:spectrum}
E(k) \equiv \left\langle
\frac{1}{2} \int 
d\hat{\vec{k}}
\ |\fvec{u}(\vec{k})|^2  
\right\rangle_t \ ,
\eeq
with $\hat{\vec{k}}=\vec{k}/|\vec{k}|$ a unit vector, 
is the time-averaged energy 
spectrum after reaching a statistically steady state. 
$E_1$ is shown as a function of 
the ratio $|\nu_1/\nu_0|$ in Fig.~\ref{fig:E1_vs_nu1_caseA}, 
together with typical examples of velocity fields.
At low values $|\nu_1/\nu_0| \leqslant 2$ the
energy at the largest scale 
is negligible and the corresponding 
flows at $|\nu_1/\nu_0| =2$ and  $|\nu_1/\nu_0| =1$ do not show
any large-scale structure. At a critical 
value $|\nu_{1,\rm crit}/\nu_0| = 2.06 \pm 0.02$ a sharp transition occurs so that 
for larger values of $|\nu_1/\nu_0|$ a condensate 
consisting of two counter-rotating vortices  
at the largest scales exists (see the case $|\nu_1/\nu_0| =5$
in Fig.~\ref{fig:E1_vs_nu1_caseA}, and Refs.~\cite{Chertkov07,Xia09,Laurie14}).
Since a condensate can only build up once the transfer of kinetic energy
reaches up to the largest scales, the presence of a condensate is a tell-tale
sign of an inverse energy transfer. 

For $|\nu_1|\gg|\nu_{1,\rm crit}|$, we observe $E_1 \sim \nu_1^2$, which can be
rationalized by mapping the large-scale dynamics onto an Ornstein-Uhlenbeck
process \cite{Ornstein30,Chandrasekhar43,Risken96,vanKampen07}. 
Neglecting small-scale dissipation, Eq.~\eqref{eq:pcv-spectral} can
formally be written as $\partial_t \omega = -\nu_0\Delta\omega_{\rm LS} -
\nu_1\Delta\omega_{\rm IN}$, where $\omega_{\rm LS}$ and $\omega_{\rm IN}$ are
the vorticity field fluctuations at scales larger and smaller than $\pi/k_{\rm
min}$, respectively.  For $\omega_{\rm LS}$, this results in an
Ornstein-Uhlenbeck process with relaxation time $1/\nu_0$ and diffusion
coefficient $\nu_1^2/2$, because $\omega_{\rm IN}$ can be considered as noise
on the time-scale of $\omega_{\rm LS}$.  Therefore, $E_1 \simeq E_{\rm LS} \sim
\nu_1^2/\nu_0$.

\begin{figure}[tbp]
\centering
	\includegraphics[width=\columnwidth]{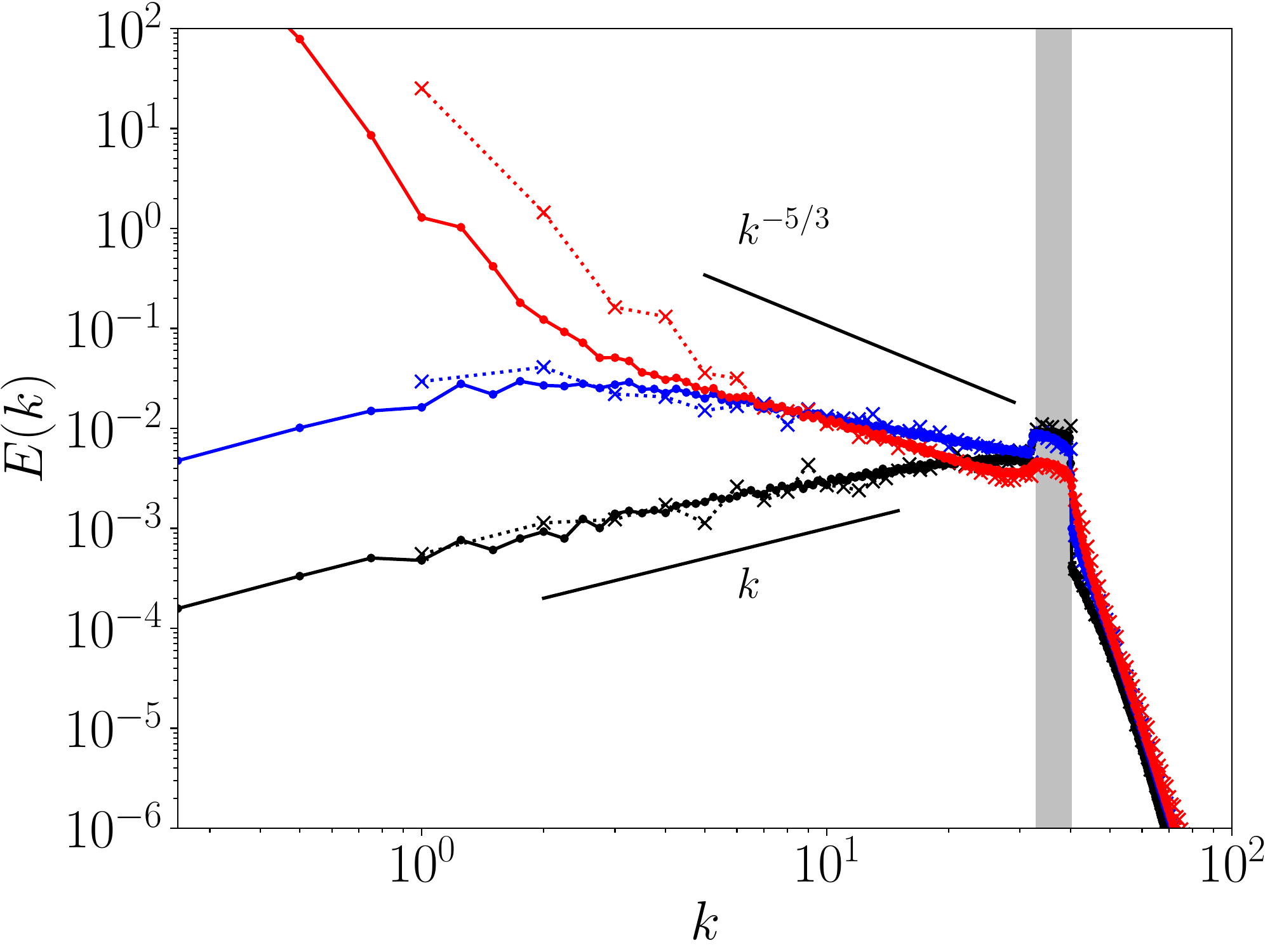}\\
	\includegraphics[width=\columnwidth]{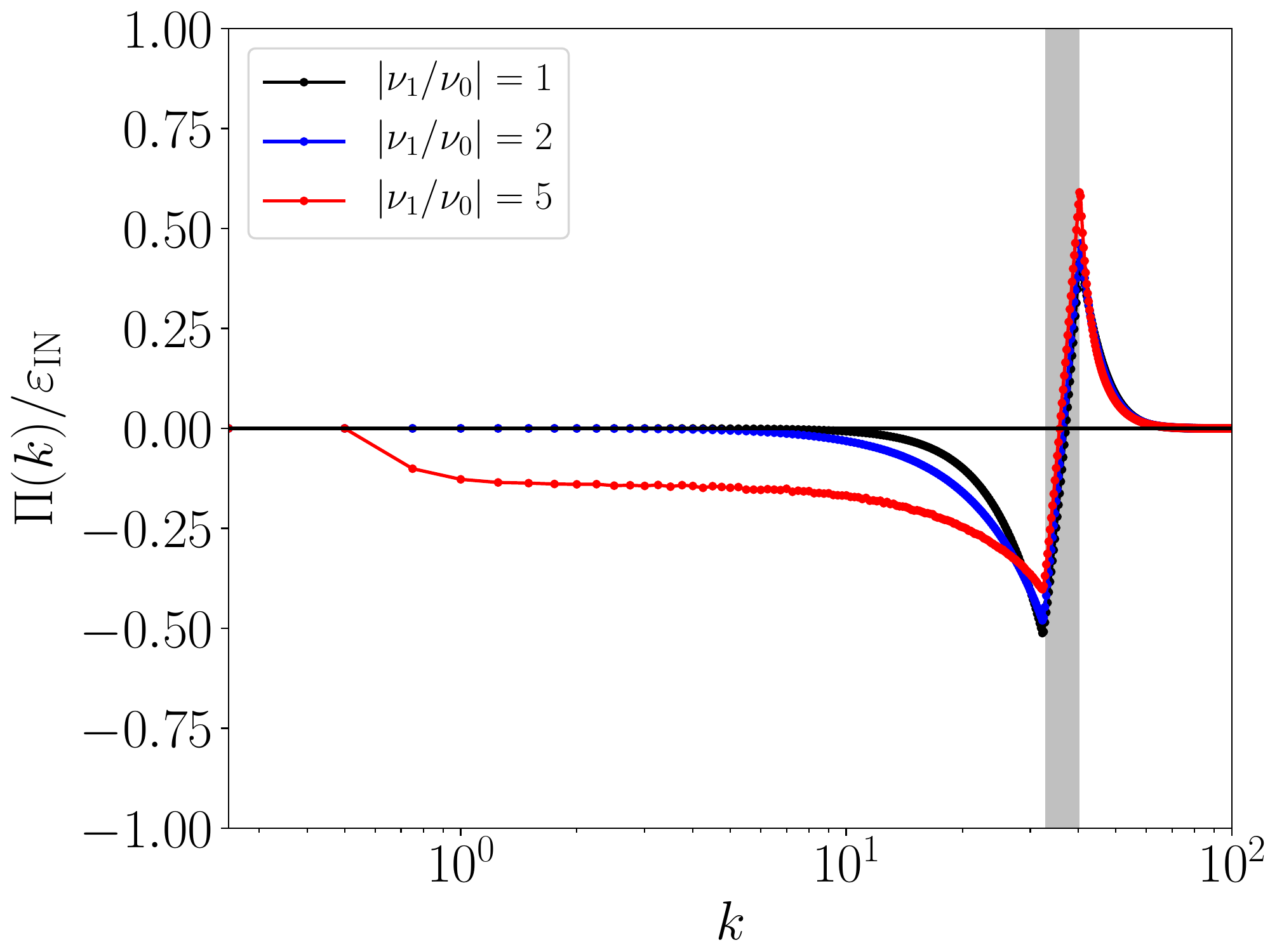}
 \caption{
(Color online) 
Energy spectra (top) and fluxes (bottom) for three example cases 
$|\nu_1/\nu_0| = 1$ (black) , $|\nu_1/\nu_0| = 2$ (blue) and $|\nu_1/\nu_0| = 5$ (red) 
for $N=256$ (dotted lines) and $N=1024$ (solid lines). The higher-resolved data has been rescaled 
according to Eq.~\eqref{eq:scaling} to account for $k \to 4k$. 
The gray-shaded area indicates the interval $[k_{\rm min},k_{\rm max}]$, and 
the solid lines in the top panel correspond to theoretical predictions, i.e.
energy equipartion: $E(k) \sim k$, and Kolmogorov scaling: $E(k) \sim k^{-5/3}$.
         }
 \label{fig:spectra_fluxes_caseAC}
\end{figure}

The transition and its precursors can be analyzed in terms of energy spectra,
shown in the top panel of Fig.~\ref{fig:spectra_fluxes_caseAC} for three
typical examples.  As expected from the large-scale pattern observed for the
case $|\nu_1/\nu_0| = 5$, the corresponding energy spectrum shows the
condensate as a high energy density at $k=1$. In the other two cases,
$|\nu_1/\nu_0| = 1$ and $|\nu_1/\nu_0|=2$, the energy density tapers off
towards small wave numbers, and there is no condensate.  The spectra for $k\le
k_{\rm min}$ follow power laws, with exponents in the range set by energy equipartion
where $E(k) \sim k$, and a Kolmogorov scaling, $E(k) \sim k^{-5/3}$, as
indicated by the solid lines in the figure.  
\blue{
The spectral exponent is known to depend on large-scale dissipation, if present
\cite{Bratanov15}, and on the presence of a condensate \cite{Chertkov07}.
}
For the case $|\nu_1/\nu_0| = 1$,
the energy spectrum is $E(k) \sim k^{0.75}$, and close to the equipartition
case.  With increasing amplification factor the spectral exponent turns
negative, with $E(k) \sim k^{-0.75}$ for $|\nu_1/\nu_0| = 2$ and $E(k) \sim
k^{-1.2}$ for $|\nu_1/\nu_0| = 5$.    

The occurrence of states close to absolute equilibrium in the region $k<k_{\rm
min}$ for weak forcing suggests the presence of a second transition to a net
inverse energy transfer for stronger forcing, as in the case $|\nu_1/\nu_0| =
2$.  Although the spectral exponent in this case suggests that energy is
transferred upscale, the absence of a condensate implies that this energy
transfer must stop before reaching $k=1$.  The flux of energy across scale $k$
in the statistically steady state can be measured with
\beq
\label{eq:flux}
\Pi(k)  \equiv - \left \langle  \int_{|\vec{k}'|\leq k}
d\vec{k}'  \ \fvec{u}(-\vec{k}') \cdot\mathcal{F}_{\vec{k}'}{\left[(\vec{u} \cdot \nabla)\vec{u} \right]} \right \rangle_t .
 \
\hspace{-0.5em}
\eeq
The sign of $\Pi(k)$ is defined such that $\Pi(k) < 0$ 
corresponds to an inverse energy transfer
and $\Pi(k) > 0$ to a direct energy transfer. 
As shown in Fig.~\ref{fig:spectra_fluxes_caseAC}, bottom panel,
the fluxes tend to zero as $k$ tends to $1$ for
$|\nu_1/\nu_0| = 1$ and $|\nu_1/\nu_0| = 2$, 
indicating that the inverse energy transfer is 
suppressed 
by viscous dissipation close to $k_{\rm min}$. 
In contrast, for $|\nu_1/\nu_0| = 5$,  the flux $\Pi(k) \simeq \rm const. $,
clearly indicating an inertial range 
and hence an inverse energy cascade in the strict sense,
as expected for a hydrodynamic energy transfer that is 
dominated by the \blue{inertial} 
term in the Navier-Stokes equations.

\begin{figure}[tbp]
\vspace{-1.5em}
\centering
	\includegraphics[width=\columnwidth]{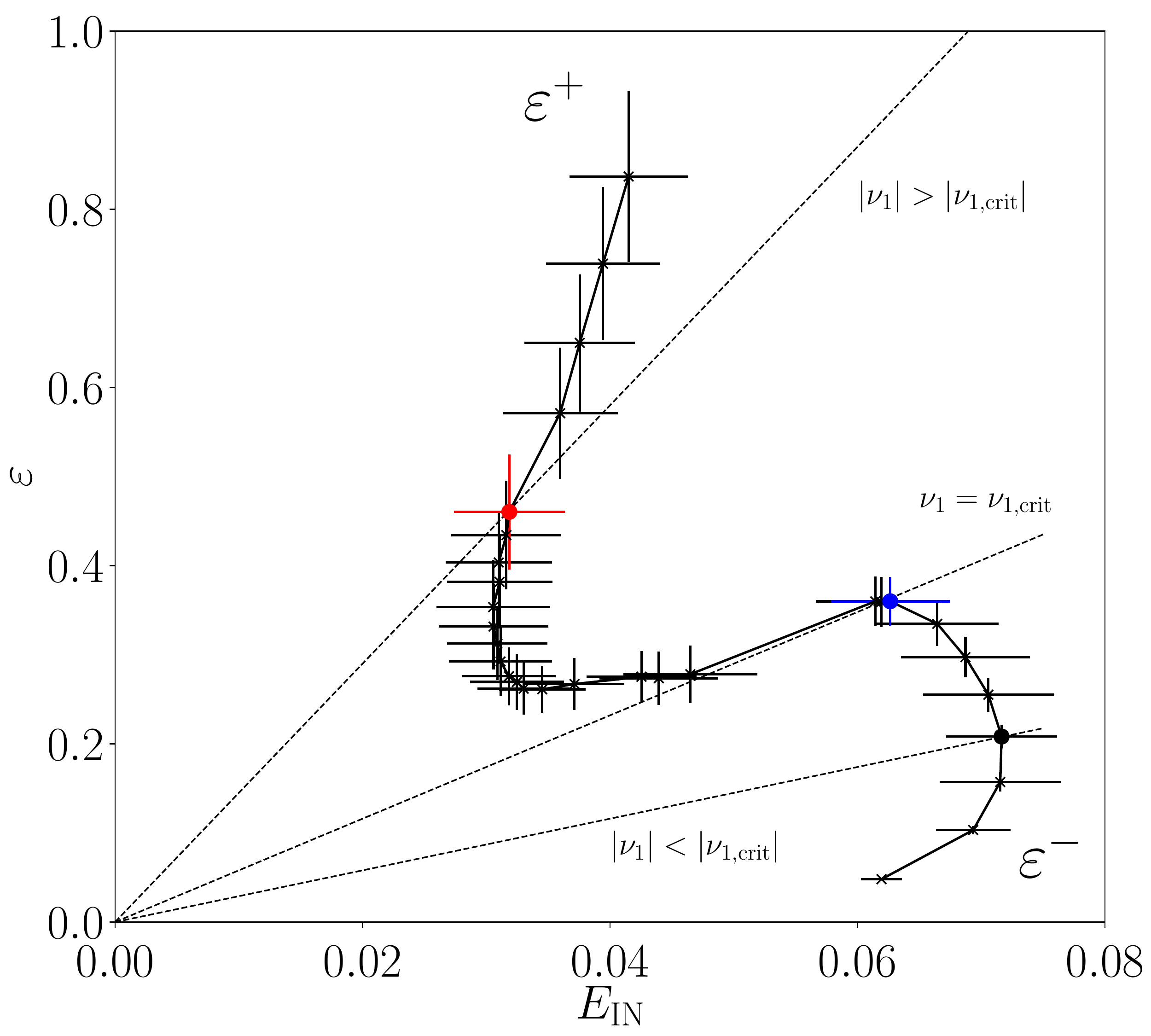}
 \caption{
         (Color online) Total mean dissipation rate as a function of the energy in the 
         interval $[k_{\rm min}, k_{\rm max}]$.
         The dashed lines indicate different values of the amplification factor. 
         Error bars indicate the standard deviation.
         The larger open symbols in black, blue and red 
         correspond to cases 
         $|\nu_1/\nu_0| = 1$, $|\nu_1/\nu_0| = 2$ and $|\nu_1/\nu_0| = 5$, respectively.
         }
 \label{fig:DT_vs_Einput}
\vspace{-1em}
\end{figure}

The transition shows up not only in the energy transfer across scales, but also
in the total energy balance. The special form of Eq.~(\ref{eq:momentum}) with the
piecewise constant viscosity as in Eq.~(\ref{eq:nu_k}) gives a balance between the energy
contained in the forced modes, $\Ein \equiv \int_{k_{\rm min}}^{k_{\rm max}} dk
\ E(k)$, and the dissipation in the other wave number regions, $\varepsilon  =
2\nu_0\int_0^{k_{\rm min}} dk \ k^2E(k) + 2\nu_2 \int_{k_{\rm max}}^\infty dk \
k^2 E(k)$.  In a statistically stationary state $\varepsilon \simeq 2k_{\rm
f}^2 |\nu_1| \Ein$, where $k_{\rm f}=(k_{\rm min} + k_{\rm max})/2$ corresponds
to an effective driving scale.  Figure \ref{fig:DT_vs_Einput} presents the
relation between $\varepsilon$ and $\Ein$, obtained from simulations for
different $\nu_1$. Statistically stationary states are obtained as crossings
between $\varepsilon (\Ein)$ (the symbols connected by continuous lines) and
the equilibrium condition $\varepsilon \simeq 2k_{\rm f}^2 |\nu_1| \Ein$, shown
by dashed lines for different $\nu_1$. 

For small $|\nu_1|$ the energy content in the forced 
wave number range increases with $|\nu_1|$. However, as the transfer to 
a wider range of wave numbers sets in dissipation increases, and the energy
$\Ein$ decreases (branch labelled $\eps^-$). 
This is a smooth transition from absolute equilibrium 
to viscously damped inverse energy transfer.
At the critical forcing $|\nu_{1,crit}|$, both $\varepsilon$
and $\Ein$ drop, and a gap forms: the signal of the first-order
phase transition. Further increasing $|\nu_1|$ results in even lower $\Ein$,
with only small variations in $\varepsilon$, so that
$\Ein \sim |\nu_1|^{-1}$. In this region, the
dynamics cannot be dominated by the condensate. 
Eventually, the condensate 
takes over the energy dissipation; the curve turns around to give 
$\eps \propto E_1 \propto |\nu_1|^2$ and $\Ein \propto \eps^{1/2}$
(branch labelled $\eps^+$). 
In this region, a strong condensate will alter the
nonlinear dynamics \cite{Chertkov07,Laurie14} and the characteristic 
Kolmogorov scaling of $E(k)$ for 2d turbulence disappears. 
Finally, the particular S-shape of the curve shows that two 
non-equilibrium steady states corresponding to the branches $\varepsilon^+$ and $\varepsilon^-$, 
respectively, can be realised for the same value of the energy $\Ein$
in the forced range.
The existence of two stable branches connected by an unstable region     
describes the bistable scenario characteristic of a first-order 
non-equilibrium phase transition.

In order to relate the numerical data to experimental results, we now compare
the Reynolds numbers and characteristic scales involved in active suspensions
and in our simulations.  
For a suspension of {\it B. subtilis}, the
characteristic size of the generated vortices is about $100 \mu m$ with a
characteristic velocity around $35-100 \mu m$  \cite{Dombrowski04}, resulting
in ${\rm Re}_{\rm vortex} = O(10^{-2}-10^{-3})$.  Taking into account a
possible reduction in viscosity down to a `superfluid' regime measured
experimentally for 
{\it \blue{Escherichia} coli}
\cite{Lopez15}, a Reynolds number regime of
$O(10)$ seems possible, provided the density of the suspension is not too high.
Larger microswimmers may lead to even higher Reynolds numbers, 
with values of around 30 for magnetic spinners accompanied by 
Kolmogorov scaling of $E(k)$ \cite{Kokot2017}, 
and 25 for camphor boats (C.~Cottin-Bizonne, private communication). 


The forcing in our equations models the scale of such vortices, so we need a corresponding
Reynolds number for the comparison between the model and potential realizations.
With $k_{\rm f} = (k_{\rm min} + k_{\rm max})/2$ the center of the forced modes,
and $\Ein$ the energy in these modes, we can define
${\rm Re}_{\rm B} = \sqrt{\Ein}(\pi/k_{\rm f})/\nu_0$. 
Just above the critical point, we measure 
${\rm Re}_{\rm B} \simeq 15$. While these value are still 
larger than the typical Reynolds number of active suspensions, 
an experimental realization
of the transition seems within reach.

This comparison also gives relations for the length and time scales.  Setting
the forcing scale to $\pi/k_{\rm f}=50\mu m$, the lattice with $256^2$
collocation points corresponds to a box length of $3600\mu m$, larger than the
usual experimental domain sizes. It is possible to detect the formation of the
vortices also in smaller domains, but then it will be difficult to extract
scaling exponents for the energy densities and the energy flux.  For the time
scales, the comparison is more favorable, with the large-eddy turnover time $T$
and the characteristic timescale of the mesoscale vortices $L_{\rm
box}/\sqrt{\Ein}$ resulting in $0.4s \leqslant T \leqslant 0.8s$, and hence a
run time of $20-40min$ for the different simulations. For comparison, constant
levels of activity in  {\it E. coli} can be maintained for several hours
\cite{Lopez15}. 

The systematic parameter study of a hydrodynamic model applicable to dense
suspensions of microswimmers presented here shows a sharp transition between
spatio-temporal chaos (bacterial turbulence) and large-scale coherent
structures (hydrodynamic turbulence). The transition is preceded by a
statistically steady state  in which a net inverse energy transfer is damped by
viscous dissipation at intermediate scales before reaching the largest scales
in the system. Above the critical point, a condensate forms at the largest
scales and the energy flux is scale-independent over a range of scales, i.e.,
the flows in that parameter range are hydrodynamically turbulent.  A comparison
between the driving-scale Reynolds number in our simulations and typical
Reynolds numbers of active suspensions suggests that it should be possible to
observe the transition to large scale coherent structures also experimentally.
Our results should be generic for active systems
where the forcing is due to linear amplification. For instance,
we verified that also in the continuum model (Eq.~\eqref{eq:stress_fourier}) 
the condensate forms suddenly under small changes in forcing
at similar Reynolds numbers as in the PCV model
\blue{\cite{Linkmann19b}}. 

Finally, we note that in rotating Newtonian fluids, transitions to condensate
states have been observed as a function of the rotation rate (Rossby number)
\cite{Alexakis15,Yokoyama17,Seshasayanan18}.  This suggests that the appearance
of a condensate may be connected with a phase transition also in other flows.

GB acknowledges financial support by the Departments of Excellence Grant
(MIUR).  MCM was supported by the National Science Foundation through Grant
DMR-1609208. 

\bibliography{references}

\end{document}